# Probabilistic Models For Joint Clustering And Time-Warping Of Multidimensional Curves

**Darya Chudova, Scott Gaffney, and Padhraic Smyth**
School of Information and Computer Science
University of California, Irvine
Irvine CA 92697-3425
{dchudova,sgaffney,smyth}@ics.uci.edu

## Abstract

In this paper we present a family of models and learning algorithms that can simultaneously align and cluster sets of multidimensional curves measured on a discrete time grid. Our approach is based on a generative mixture model that allows both local nonlinear time warping and global linear shifts of the observed curves in both time and measurement spaces relative to the mean curves within the clusters. The resulting model can be viewed as a form of Bayesian network with a special temporal structure. The Expectation-Maximization (EM) algorithm is used to simultaneously recover both the curve models for each cluster, and the most likely alignments and cluster membership for each curve. We evaluate the methodology on two real-world data sets, and show that the Bayesian network models provide systematic improvements in predictive power over more conventional clustering approaches.

## 1 Introduction and Motivation

Data in the form of sets of curves arise naturally across a variety of applications in science, engineering, and medicine. Examples include time-course measurements from sets of genes (Eisen et al., 1998), estimated trajectories of individuals or objects from sequences of images (Gaffney & Smyth, 1999), and biomedical measurements of the response of different individuals to drug therapy over time (James & Sugar, to appear, 2003). In this paper we will use the term "curve" to denote a variable-length series of data measurements observed as a function of some independent variable such as time. More generally, each individual "curve" can consist of set of multi-dimensional (vector) measurements as a function of the independent variable, rather than a single measurement. For example, in modeling spatial trajectories of objects the measurements at each time point can include the spatial $x, y$ location of the object as well as features characterizing the object's state (shape, motion parameters, and so forth). In statistics, such data are sometimes referred to as "functional data" (Ramsay & Silverman, 1997), emphasizing the fact that the observed data are functions of an independent variable.

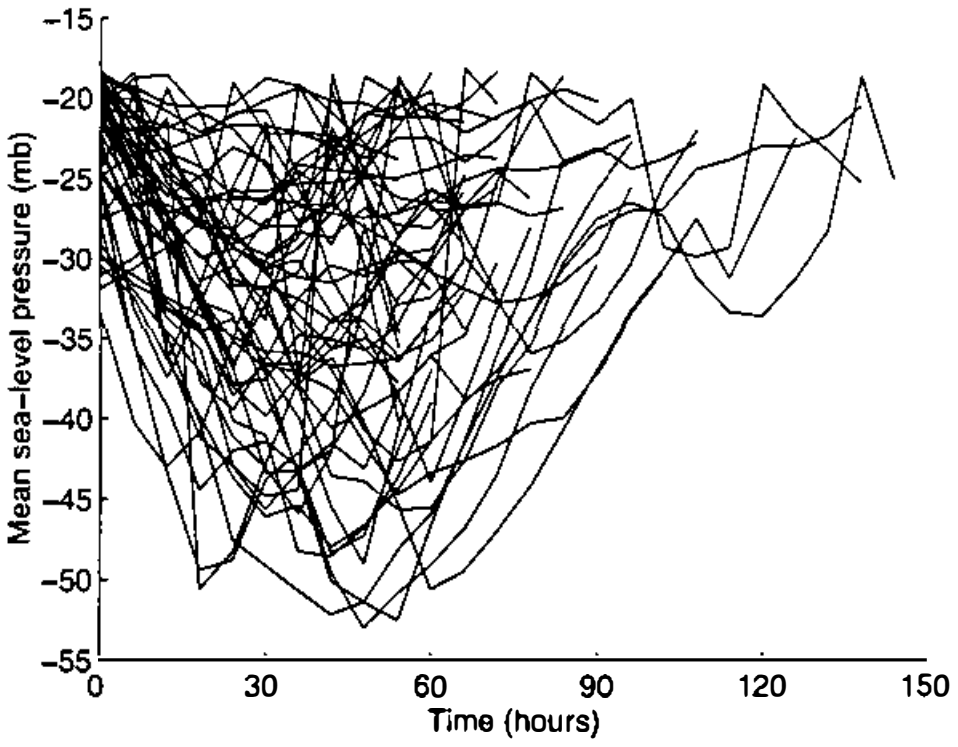

Figure 1: Cyclone intensities as a function of time.

As an example, Figure 1 shows a set of 1-dimensional curve data where each individual curve represents the intensity of the center of a particular cyclone over its lifetime. Figure 2 shows another set of curves, this time a set of heartbeats extracted from an ECG time-series. A practical problem with such data is that the curves tend to be misaligned in various ways. For example, in both Figures 1 and 2, variations among curves in the amplitude and time axes could be substantially reduced by aligning individual curves in different ways. The lack of alignment here is both an artifact of the methods used to extract these curves (e.g., in detecting and tracking cyclone centers in sea-level pressure data, Gaffney & Smyth, 2003), as well as being due to natural variability in the underlying (and unknown) dynamic processes generating the data.

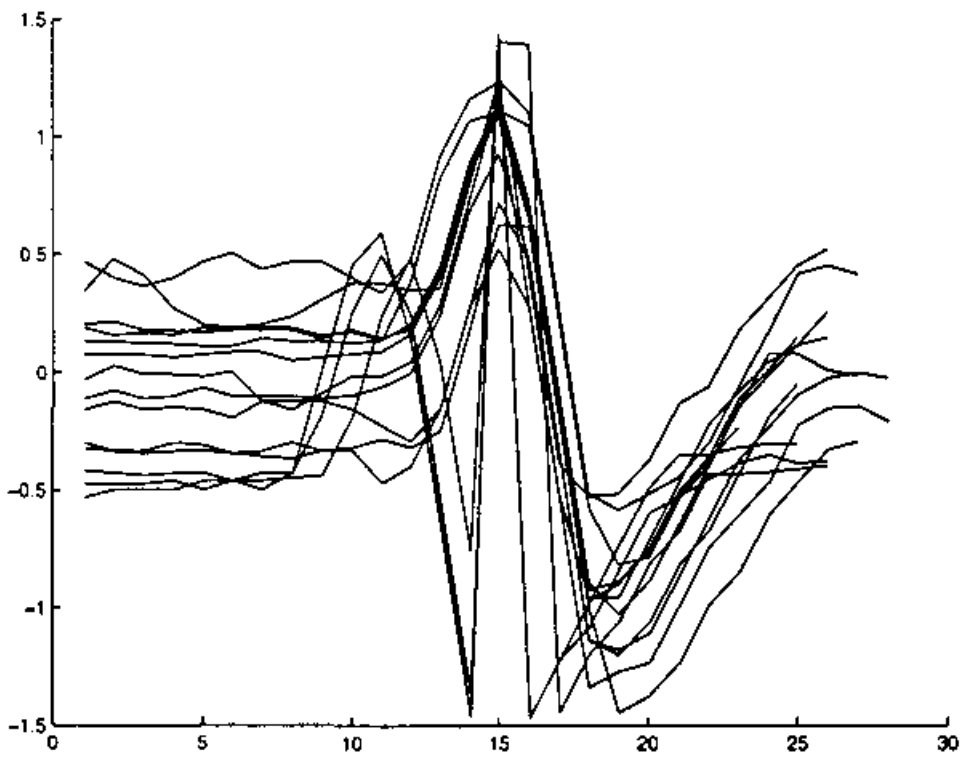

Figure 2: ECG time series segments containing heartbeats.

More generally, all manners of (unknown) transformations may have been applied to observed curve data, such as offsets and scaling in the observed measurements or more complex forms of non-linear warping of the signal. Thus, a significant body of work exists on techniques that attempt to remove various classes of transformations from observed curve data, such as time warping (Wang & Gasser, 1997), curve registration (Ramsay & Li, 1998), structural averaging (Kneip & Gasser, 1992), and point-set matching for image registration (Gold et al., 1998).

A second body of work in the analysis of sets of curves consists of curve clustering techniques. Since the data are usually being generated by a set of different individual "dynamical systems" (or the same system at different times) it is natural to ask the question in many applications as to whether different natural clusters (or regimes) of these systems exist in the data, e.g., clusters of genes, storms, or individuals. Probabilistic mixture models have been found to be particularly useful in this context. For example, regression-based mixture modeling focuses on the finding of two or more underlying functions (e.g., polynomials) from which the observed data might have been generated. This technique, known as *regression mixtures* (DeSarbo & Cron, 1988; Jones & McLachlan, 1992) extends the standard finite mixtures framework to the case where component density models are replaced with conditional regression density models. More recent work along this line focuses on learning individual models for each curve during the clustering. This can be handled for example, through the integration of linear random effects models with regression mixtures (Lenk & DeSarbo, 2000; Gaffney & Smyth, 2003). Further extensions have been developed that use non-parametric models for the mean curves, such as kernel regression models (Gaffney & Smyth, 1999) and splines (James & Sugar, to appear, 2003; Bar-Joseph et al., 2002).

In prior work to date on curve modeling, curve alignment techniques and curve clustering have largely been pursued separately and independently. Clustering in the presence of curve transformations is conceptually problematic: we may not be able to effectively cluster the data without first removing the transformations, but on the other hand we may not be able to effectively remove the transformations if we do not know the cluster memberships. One approach is to preprocess or post-process the sets of curves by employing alignment or registration techniques such as dynamic time-warping before or after clustering (Wang & Gasser, 1997). The disadvantage of such an approach is that the discovery of the curve transformations and curve clustering are decoupled from each other, which can in principle weaken the ability of the algorithm to detect structure in the data.

One area where there has been some success to date in simultaneous alignment and clustering is with image data. In particular, Frey and Jojic (2003) use EM to learn mixtures of image patches subject to linear offsets and rotation. Because 1-dimensional curve data is much more constrained than 2-dimensional pixel images (given the implicit ordering on the data points in 1-dimension) we are able to address in this paper a more general class of transformations for clustering curve data than is feasible in the 2-dimensional case, in particular, non-linear warping of the independent variable ("time-warping").

In this work we make the specific assumption that we can achieve useful results by restricting attention to transformations that are "on-grid" in terms of the independent variable, i.e., that shifts and warps are constrained to occur on the same sampling grid that the data are measured on. This is in direct contrast to "off-grid" methods that interpolate between the gridded observations, such as polynomial or spline models. The advantages of the on-grid approach (as we will see later in the paper) are that (a) we can use a completely non-parametric model for the mean curves within each cluster by avoiding parametric assumptions on the interpolating function, and (b) we get a computationally feasible procedure for solving the joint clustering/transformation problem by using a discrete-time (or discrete-grid) Bayesian network. Of course for certain applications (for example when data are very sparse for each curve) the interpolative (or functional modeling) methods might be more appropriate. In this paper, however, the focus is on the "on-grid" class of modeling techniques. Experimental results later in the paper bear out that substantial and systematic improvements in modeling power can be gained by the "on-grid" approach alone.

The primary novel contribution of this paper is the learning of clusters of curves in the presence of certain classes of curve transformations. While there has been a significant amount of prior work on each of curve-clustering and curve-transformations in isolation (as discussed above) there has been no work that we are aware of that specifically addresses simultaneous learning of clusters and transformations of curve data. The specific class of curve transformations we address includes both global discrete-valued translations and local non-linear warping along the time (or independent variable) axis, and real-valued additive offsets in the measurement (dependent variable) axes. Extensions to include other forms of transformations such as multiplicative scaling of the curve measurements can also be developed but are not specifically addressed in the paper. In a related paper (Chudova et al., 2003), we investigate curve clustering techniques that allow global offsets in both the measurement variables and independent variables, but without any local time-warping.

The paper is organized as follows. Section 2 introduces the Bayesian network model for curve clustering. Section 3 describes a general parameter estimation framework based on the EM algorithm, including a discussion of computational complexity. Section 4 describes the evaluation of the models and parameter estimation algorithms using two real datasets: clustering ECG data and clustering of extra-tropical cyclones. In Section 5 we discuss future directions and conclusions.

## 2 Probabilistic Curve Clustering

### 2.1 A Generative Model for Sets of Curves

In this section, we describe a generative model for multidimensional curves observed on a fixed time grid (henceforth in the paper we will refer to the independent variable as "time" although in general this need not be the case). The model allows for (a) heterogeneity via clustering, (b) both global and local discrete-valued shifts of the measurements in time, and (c) global real-valued translations of the measurement axes. In what follows we introduce each of these features in turn. We use boldface symbols for vectors and regular symbols for scalars.

#### 2.1.1 Heterogeneity in the observed curves

We use a finite mixture model with $K$ components to allow for heterogeneity in the generated curves: the probability of an individual curve $\mathbf{Y}$ given a set of model parameters $\Theta$ is defined as:

$$P(\mathbf{Y}|\Theta) = \sum_{k=1}^{K} \alpha_k P(\mathbf{Y}|Z=k) \quad (1)$$

where $\alpha_k$ is the probability of component $k$ and $Z$ is a random variable indicating cluster membership for curve $\mathbf{Y}$.

#### 2.1.2 Global time shifts and local time-warping

To generate curves allowing both global linear time-shifting and local non-linear time-warping, we allow the model to traverse different paths on the time grid when generating a curve from a particular component. We assign probabilities to the paths and assume first-order Markov dependence between the consecutive points on the path along the time grid. We use $\mathbf{G} = (g_1, \ldots, g_L)$ to denote a path on the time grid for generating an observation vector $\mathbf{Y}$ of length $L$, and obtain the likelihood of $\mathbf{Y}$ by marginalizing over all possible paths:

$$P(\mathbf{Y}|\Theta) = \sum_{k=1}^{K} \sum_{\mathbf{G}} \alpha_k P(\mathbf{Y}|k, \mathbf{G}) P(\mathbf{G}|k)$$

$$P(\mathbf{G}|k) = P(g_1|k) \prod_{j=2}^{L} P(g_j|k, g_{j-1})$$

To generate curves with global time shifts, we allow a path to start in any of the first $M$ points on the grid. This constraint is encoded by the initial state distribution in component $k$: $P(g_1|k)$. In the absence of any local transformations, the path then moves forward from its starting point "linearly" by one time point at a time. To produce local non-linear time deformations, we allow the path to skip up to $S$ points forward on the time grid, or to remain at the current time point with some small probability. This set of constrains is encoded by the transition distribution for component $k$: $P(g_j|k, g_{j-1})$

Figure 3 shows an example of a transition structure for a model with 3 mixture components and global (but not local) time-shifts. There are 3 possible initial starting positions on the time grid, allowing curves of lengths between 1 and 6. The circles in the diagram represent positions on the time grid within each mixture component and the arrows correspond to non-zero transition probabilities. Figure 4 shows an example of a transition structure for a model that additionally allows local skips and repeats along a path.

Note that if the observed measurements at time $j$ are conditionally independent of each other given the parameters of the model and cluster membership, the mixture model described above can be viewed as a form of hidden Markov model (HMM). The hidden states $H$ of this HMM encode both cluster member-ship and location on the time grid and correspond to the circles in Figures 3 and 4. Casting the problem in an HMM framework provides for effective inference with the proposed model. However, as we will see below, special care is required when we introduce global offsets in the *measurement axes* as the standard independence assumptions of the HMM are then violated.

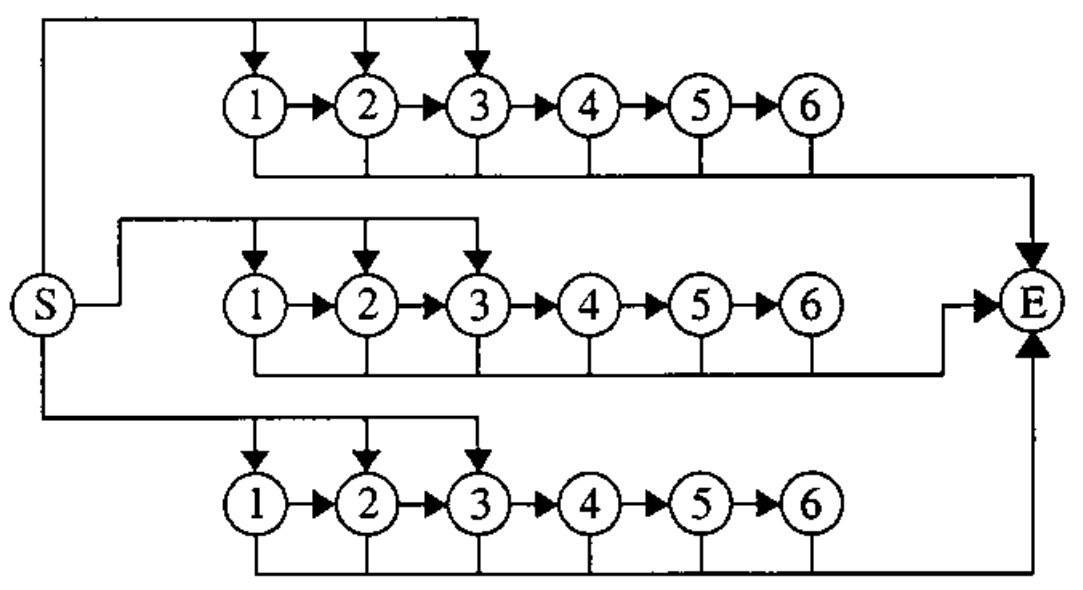

Figure 3: Transition structure of the HMM corresponding to a mixture model with linear time shifts, $K = 3$, $M$=3.

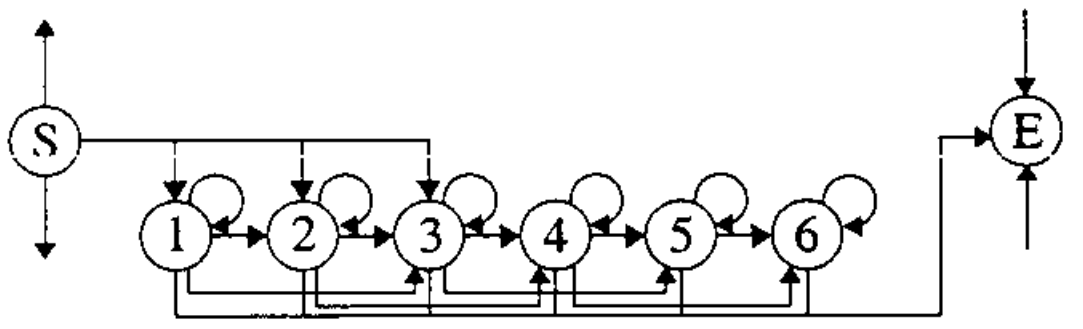

Figure 4: Transitions within a single component that allows local time-warping, $M = 3$, $S = 1$.

### 2.1.3 Translation of the measurement axes

To model the *shapes* of the curves rather than their absolute values, the generative process is invariant to the translation of the observed curves in the measurement space. To simulate curves with translations, we introduce an additional offset variable $\boldsymbol{\delta}$ per curve, and allow each mixture component to generate curves with offsets relative to the component's mean vector. The conditional likelihood of the observed curves is independent of the value of the offset, and thus curves with the same shape but different origins are identical under this generative model.

Figure 5 shows the Bayesian network structure for the proposed generative model. The network is instantiated given a particular curve with observed measurements $\mathbf{Y}$ of length $L$. The network includes the following variables: the path along the time grid $G_1, \ldots, G_L$ (unobserved), the component identity $Z$ (unobserved), the offset vector $\boldsymbol{\delta}$ (unobserved), and the measured curve $\mathbf{Y}_1, \ldots, \mathbf{Y}_L$ (observed). Naturally, we have an arrow from the component identity $Z$ to $\mathbf{Y}$'s and $G$'s, from time grid positions $G_j$ to observations $\mathbf{Y}_j$, and from every time grid position $G_j$ to the next one $G_{j+1}$, the last two as in a regular HMM. However, the observations are *not* independent given cluster membership and location on the time grid, as is the case in an HMM, but coupled through the offset vector. In relating Figure 5 and Figure 4 note that Figure 4 is a state transition diagram with edges that correspond to non-zero entries in the conditional probability table relating nodes $G_j$ and $G_{j+1}$ in Figure 5.

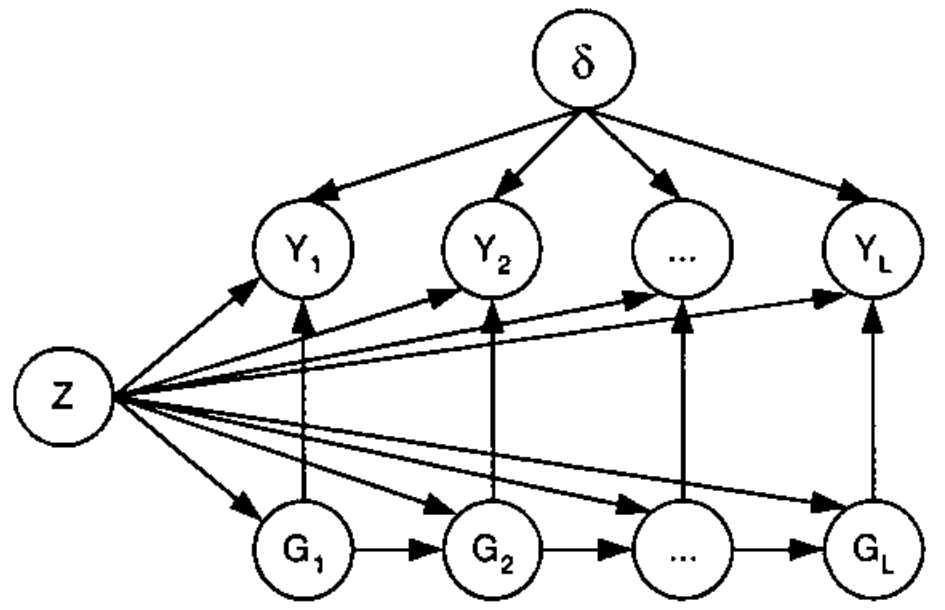

Figure 5: A Bayesian network for a curve $\mathbf{Y}$ of length $L$; measurement offsets modeled with a random variable.

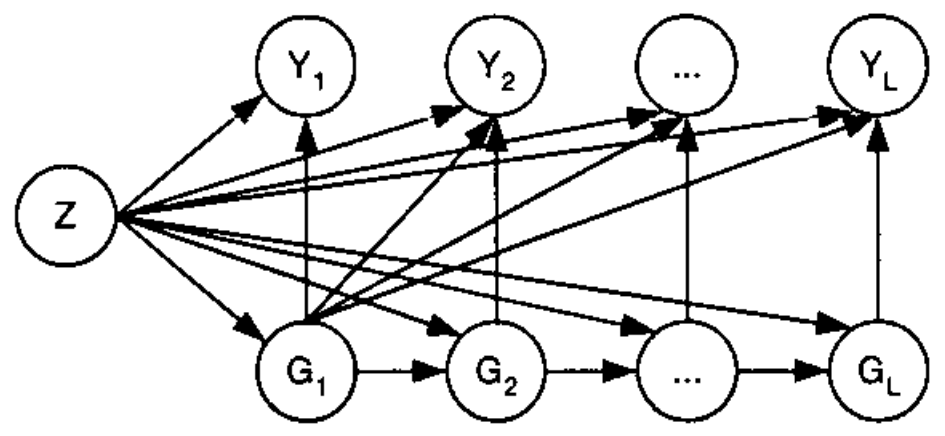

Figure 6: A Bayesian network for a curve $\mathbf{Y}$ of length $L$; measurement offsets modeled with a deterministic function.

The inference in the model pictured in Figure 5 requires summing out all possible components and paths, and integrating over the values of the offset variable. The coupling of the observed measurements through $\boldsymbol{\delta}$ prohibits the recursive computation that effectively handles exponentially large number of paths in a regular HMM, and makes the problem intractable.

To make the inference tractable we use the following heuristic approximation: we define the offset $\boldsymbol{\delta}(\mathbf{Y}, k, g_1)$ to be a deterministic function of the component identity $k$ and the assumed global time shift $g_1$. The value of the $D$-dimensional offset vector $\boldsymbol{\delta}(\mathbf{Y}, k, g_1)$ is chosen so that the translated curve $\mathbf{Y} - \boldsymbol{\delta}(\mathbf{Y}, k, \mathbf{G})$ is best aligned (under some norm) with the portion of the $k^{th}$ component's mean curve starting at position $g_1$ on the time grid. We ignore possible time skips and repeats along the path on the time grid $\mathbf{G}$ when calculating the value of the offset vector, which makes the problem tractable. This leads to a different structure for the underlying Bayesian net-work as shown in Figure 6.

We have essentially replaced the coupling of observed values through the offset $\boldsymbol{\delta}$ with a dependence on the assumed global time shift. Given $\mathbf{Y}$, $Z$ and $g_1$, we can evaluate the probability of observing $\mathbf{Y}(j)$ at $G_j$ and perform tractable inference using standard recursive algorithms.

Note that more traditional methods of dealing with measurement axes translations (such as subtracting the mean value or first measurement, etc.) are neither applicable nor optimal in the context of unknown component identity and unknown time shifts—this is confirmed by the experimental results presented later in the paper.

### 2.1.4 Conditional likelihood given a path

To model multi-dimensional curves we assume that the measurements in different dimensions are conditionally independent given a position on the time grid and component identity. The observation at a given point $j$ on the time grid, under mixture component $k$, has a multivariate normal distribution with diagonal covariance matrix $\mathbf{C}_k(j)$ centered around a $D$-dimensional mean $\boldsymbol{\mu}_k(j)$.

Given the component identity and the first point on the path along the time grid, the conditional probability of curve $\mathbf{Y}$ is evaluated in two stages: at first, the value of the optimal offset $\boldsymbol{\delta}^*$ is calculated, and then the probability of the translated curve $\mathbf{Y} - \boldsymbol{\delta}^*$ is calculated under the $k^{th}$ component's distribution:

$$\boldsymbol{\delta}^* = \operatorname*{argmin}_{\boldsymbol{\delta}} ||\mathbf{Y} - \boldsymbol{\delta} - \boldsymbol{\mu}_k(g_1 : g_1 + L - 1)||^2 \quad (2)$$

$$P(\mathbf{Y}|k, \mathbf{G}) \sim \prod_{j=1}^{L} \mathrm{N}(\mathbf{Y}(j) - \boldsymbol{\delta}^* | \boldsymbol{\mu}_k(g_j), \mathbf{C}_k(g_j)) \quad (3)$$

where $\mathbf{Y}(j)$ is the $j^{th}$ $D$-dimensional observation within a single curve $\mathbf{Y}$, and $\boldsymbol{\mu}_k(a : b)$ is the set of curve means for component $k$ defined between positions $a$ and $b$ on the time grid. We use the Euclidean norm in expression (2), but other notions of similarity could equally well be used to define the best offset vector $\boldsymbol{\delta}$ in the measurement space, e.g., based on specific prior knowledge of the process generating the data.

## 3 Parameter Estimation

We use the EM procedure to estimate the parameters of the model described above, given a set of observed curves. The EM algorithm uses the standard message passing computations involved in estimating the parameters of a Bayesian network, where the specific network is that shown in Figure 6 with transition structure as provided in Figure 4. The network includes a number of undirected cycles due to the edges connecting the $G$'s and cluster variable $Z$ with other nodes in the network. We use loop cut-set conditioning on the values of $G_1$ and $Z$ when calculating the distributions over the other latent variables in the E-step and sufficient statistics in the M-step.

The time complexity of a single iteration of this algorithm is linear with respect to each of the following: the number of curves $N$, the dimensionality of the curves $D$, the number of clusters $K$, the maximum amount of shifting allowed $M$, and the maximum number of states reachable from any state $S + 2$. However, it is quadratic in the length of the curves $L$ (if we assume for simplicity that all curves have the same length), resulting in an overall time complexity of $O(NDKMSL^2)$. If we consider the model without local time-warping, the time complexity is linear in all remaining parameters: $O(NDKML)$. The complexity becomes quadratic in $L$ when we have to evaluate the alignments of curves to cluster centers with time skips. This is expected since the alignment time for non-probabilistic dynamic time warping algorithms is also quadratic in the length the curves.

The generative model described above treats consecutive mean values (at time $t$ and $t + 1$) for the cluster mean curves as being independent. In practice it is often reasonable to assume that the means are dependent, or equivalently that the underlying true cluster mean curves should have some degree of smoothness. In Chudova et al. (2003) we describe how the model above can be extended to include a hierarchical Bayesian prior on the means, where the first level of the hierarchy introduces dependence between the means $P(\boldsymbol{\mu}_k^{t+1}|\boldsymbol{\mu}_k^t, \sigma^2)$, and the second level controls the degree of smoothness $P(\sigma^2)$. We have found that this Bayesian smoothing produces systematically better out-of-sample predictions when the number of observed data points is relatively small. In the results reported in this paper we do not use these Bayesian smoothing priors and instead use a simpler maximum a posteriori (MAP) approach with standard Dirichlet priors on the Bayesian network probability tables and maximum likelihood estimates of the curve means. More complete details on the Bayesian estimation results are available in Chudova et al. (2003).

## 4 Experimental Results

We evaluate the proposed methodology for joint clustering and time-warping of curves using two differ-ent real-world data sets: (a) ECG time series data and (b) cyclone trajectories. To analyze the performance of the models we use cross-validated log probability (logP) scores (average log-likelihood assigned to unseen curves) and in-sample within-cluster variance. While we expect the within-cluster variance to be reduced due to time warping and alignment, it is informative to see precisely how much gain can be achieved using these techniques. The out-of-sample scores are obtained using 10-fold cross-validation. All methods are allowed 5 random starts of EM at each fold, and initialization is carried out by selecting $K$ random curves as the initial $K$ cluster means. Measurement offset vectors (the $\delta$ parameters) are estimated in all models. Due to space limitations only a subset of results are reported here—more complete experimental results, including applications to time-course gene expression data can be found in Chudova et al. (2003).

Table 1: Performance of Gaussian mixtures and models with time-shifting and warping on the ECG dataset.

| Local Time Warping | Global Time Shifting | log P | Within-Cluster StDev |
|---|---|---|---|
| None | None | 1.06 | 0.072 |
| $S = 2$ | None | 1.29 | 0.057 |
| $S = 2$ | $M = 2$ | 1.32 | 0.053 |

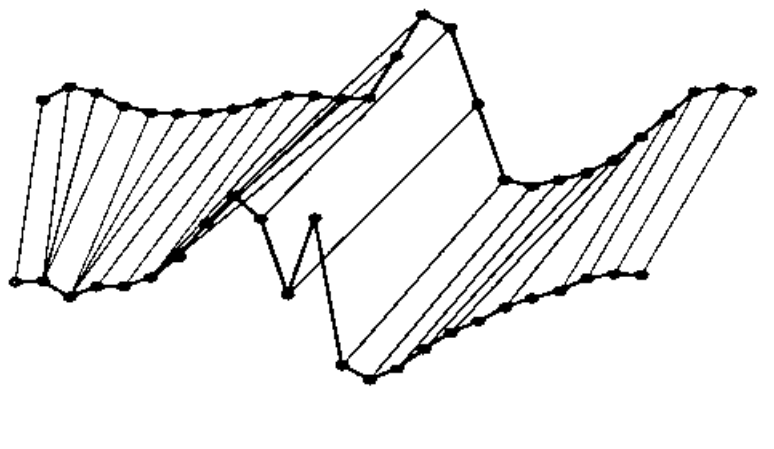

Figure 7: Sample alignment of the ECG data (top) with the recovered cluster mean (bottom).

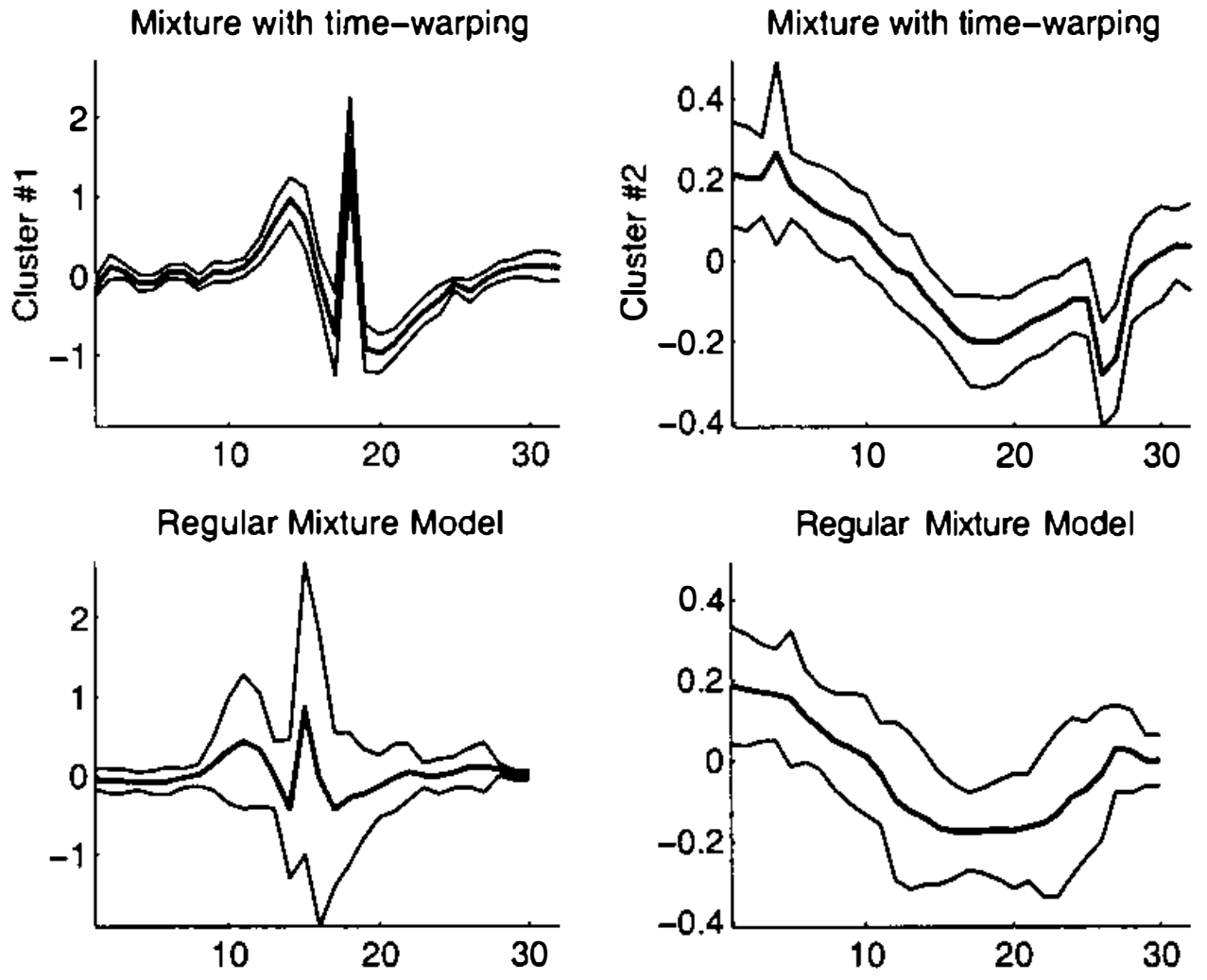

Figure 8: Cluster models with time transformations (top) and a standard mixture of Gaussians model (bottom): cluster means ± 2 standard deviations.

### 4.1 ECG Time Series

The ECG heartbeat is a clearly recognizable waveform shape—yet different examples of heartbeats even from the same individual can not necessarily be easily aligned due to the complex dynamics and variation in heartbeat generation over time (e.g., see Figure 2). We used a set of ECG data from Percival and Walden (2000). We chose a section of an ECG time-series from a particular individual, and manually divided this into 14 segments of slightly varying lengths, each corresponding to a single heartbeat cycle. We further split each segment into 5 pieces, so that each piece corresponds to a particular phase of the cycle—thus, we expect that there are 5 natural clusters in the data. The resulting 70 curves were provided as input to the joint clustering and alignment algorithm with the number of mixture components $K = 5$. Figure 2 shows data from one of the resulting clusters (without any alignment). Note that the heartbeat data we used in our experiments do not correspond to a real application (although the data itself is real). The goal is to provide an illustrative example of a broad class of "waveform recognition" problems encountered in medical and industrial applications where waveform shape is an important characteristic of the underlying dynamic process.

We report the results of applying both a standard Gaussian mixture model and models with transformations to this data set. In the standard mixture approach, we treat the curves as points in an $L_{\max}$ dimensional space, where $L_{\max}$ is the length of the longest curve. We treat the $L_{\max} - L_i$ measurements at the end of curve $\mathbf{Y}_i$ as missing (clearly it would be better to try to align curve $\mathbf{Y}_i$ in some manner, but this is intended to be a baseline technique). We also used models with global time shifts ($M = 2$) and local repeats and time-skips ($S = 2$). Table 1 contains cross-validated logP scores and within-cluster variance for different methods. Models with local time-warping and global time-shifting produce systematically higher

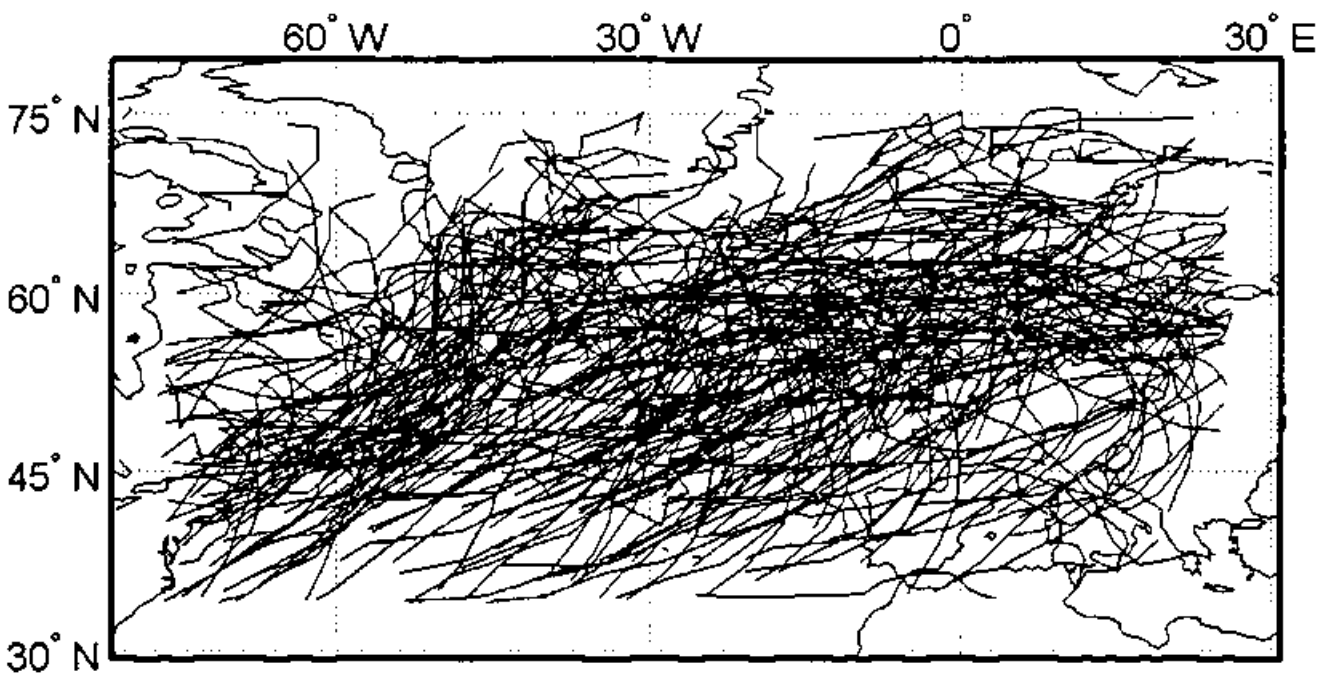

Figure 9: A sample of cyclone trajectories over the North Atlantic.

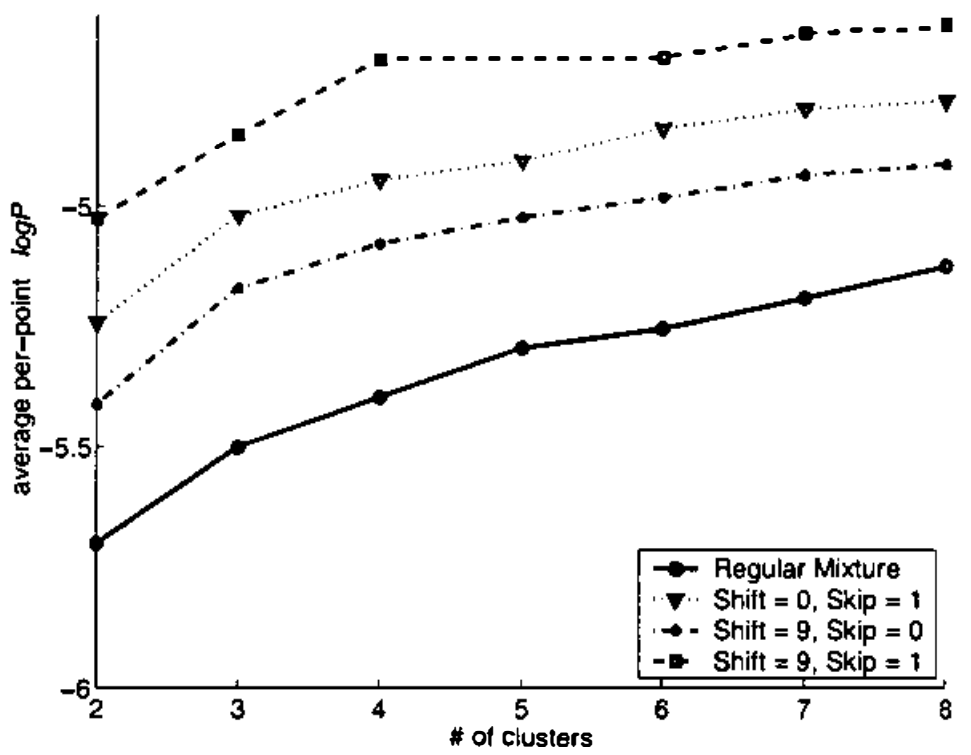

Figure 10: Cross-validated logP scores on the cyclone data.

logP scores and lead to reductions of about 25% in within-cluster variance.

Figure 7 shows a typical (most probable) alignment of an observed curve with a learned cluster mean. Figure 8 shows two of the clusters recovered by the joint clustering and alignment model (top plots), and the most similar cluster learned by a standard Gaussian mixture model from this dataset (lower plots). We see clearly that time warping greatly reduces the within-cluster variance and extracts more structure from the data, such as sharper peaks and valleys in the cluster waveforms.

### 4.2 Cyclone Clustering

We also applied our methodology to clustering of extra-tropical cyclone trajectories. The observed curve data consists of 614 multidimensional curves ($D = 3$), each curve consisting of $x, y$ lat-lon pairs and sea-level pressure intensities (as in Figure 1), corresponding to the estimated minima of the cyclones. Measurements are every six hours from the estimated beginning to the estimated end of a storm, and detected storms typically have lengths from 3 to 6 days (12 to 24 time points). Details on detection and tracking algorithms are provided in Blender, Fraedrich, and Lunkeit (1997). Figure 9 shows the superposition of a set of such trajectories, estimated from data over 15 winters. Atmospheric scientists are interested in understanding whether the cyclones can be grouped into natural clusters. The distribution of cyclone paths and intensities as a function of long-term climate variation is of particular interest, for example, understanding how the probability of extreme storms over western Europe may vary as a function of climate change. In this context, even relatively simple probabilistic modeling and clustering can be quite valuable as a means to explore the rather complex curve data shown in Figure 9 (Gaffney & Smyth, 2003). Prior work in atmospheric science on this problem has typically converted the trajectory data into a vector space (by forcing all cyclones to be of the same length in time) and then using algorithms such as K-means for clustering (Blender et al., 1997).

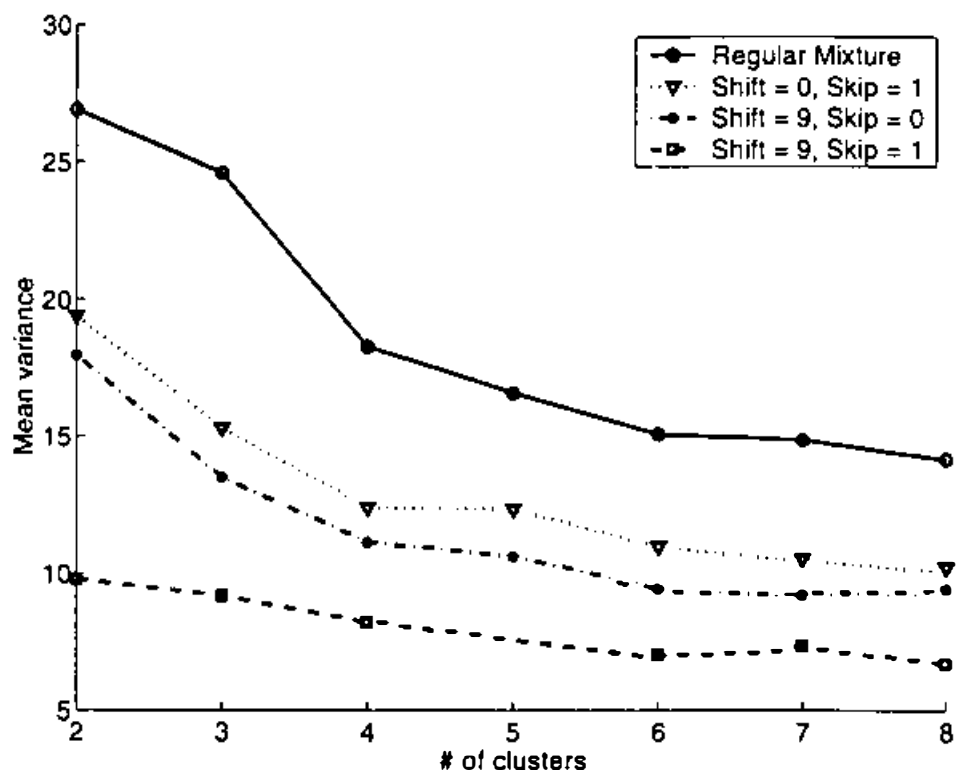

Figure 11: Mean within-cluster variance on the cyclone data.

Figure 10 shows a plot of cross validated logP scores as a function of the number of clusters $K$ for 4 different methods from our framework: (1) Gaussian mixtures as for the ECG experiments but where all cyclones are spatially shifted a priori to start at $x = 0, y = 0$, (2) a model with global shifting but no local time-warping, (3) the reverse of method (2), and (4) both global time-shifting and local time warping. The maximum shift allowed is $M = 9$ and for computational reasons we limited the maximum number of skips $S$ in the local time-warping model to 1 skip (Figure 4). We see that the more complex models (including global shifting, then local time-warping, then both) lead to systematic improvements in out-of-sample logP scores across all $K$ values. For example, the models that allow both time warping and shifts have higher scores at $K = 3$ than conventional Gaussian mixtures with $K = 8$. Figure 11 shows the same type of plot for the mean within-cluster variance. Again the model with both local time-warping and global shifts systematically produces the lowest within-cluster variance, on the order of 50% reduction compared to Gaussian mixtures (in agreement with the visual evidence in Figure 8 for the ECG data).

## 5 Conclusions

In this paper we addressed the general problem of clustering multi-dimensional curve data where we allow for local and global translations in the independent variable (typically time) and global offsets in the measurement variables. We proposed a general mixture model framework for this problem and demonstrated on two real-world data sets that the methodology systematically leads to lower variance clusters in-sample, and better predictions in terms of density estimation out-of-sample. We also argue that the methodology leads to more interpretable clusters which is often important from a scientific viewpoint. Space limitations prevented a full discussion of many other aspects of this problem. For example, it is quite easy to allow for multiplicative amplitude scaling using this same mixture framework and our experiments to date indicate that it also leads to systematically better clustering results. Bayesian estimation of the deformation parameters (e.g., via priors on the parameterized skip lengths and shifts in time) is also feasible and likely to be useful in practice.